\documentclass[aps,pre,twocolumn,superscriptaddress]{revtex4}
\usepackage{graphicx}

\begin{document}  
\title{Reply to ``Incommensurate vortices and phase transitions in two-dimensional XY models with interaction having auxiliary minima''
by S. E. Korshunov}

\author{Gabriel A. Canova}
\affiliation{Instituto de F{\'\i}sica, Universidade Federal do Rio
  Grande do Sul CP 15051, 91501-970 Porto Alegre RS, Brazil}

\author{F\'abio Poderoso}
\affiliation{Instituto de F{\'\i}sica, Universidade Federal do Rio
  Grande do Sul CP 15051, 91501-970 Porto Alegre RS, Brazil}

\author{Jeferson J. Arenzon}
\affiliation{Instituto de F{\'\i}sica, Universidade Federal do Rio
  Grande do Sul CP 15051, 91501-970 Porto Alegre RS, Brazil}
\affiliation{Universit\'e Pierre et Marie Curie-Paris VI, LPTHE UMR
  7589, 4 Place Jussieu, FR-75252 Paris Cedex 05, France}

\author{Yan Levin}
\affiliation{Instituto de F{\'\i}sica, Universidade Federal do Rio
  Grande do Sul CP 15051, 91501-970 Porto Alegre RS, Brazil}

\begin{abstract}
We present a rigorous proof and extensive numerical simulations showing the existence
of a transition between the paramagnetic and nematic phases, in
a class of generalized XY models.
This confirms the topology of the phase diagram calculated by  Poderoso et al.
[PRL 106(2011)067202]. The results disprove the heuristic argument presented by
Korshunov in arXiv:1207.2349v1, against the existence of the generalized-nematic
phase in a model with $q=3$.
\end{abstract}

\maketitle


In a recent Letter~\cite{PoArLe11}, we have studied the phase diagram 
of a generalized XY model with Hamiltonian 
$\mbox{H} = - \sum_{\langle i j \rangle} [\Delta\cos(\theta_i - \theta_j) +  (1 - \Delta) \cos(q\theta_i - q\theta_j)]$, 
where $q >1 $ is an integer and $0\leq\Delta\leq 1$. 
Using Monte Carlo (MC) simulations, we showed
that for $q=3$ and $0<\Delta\lesssim 0.4$,
the model exhibits --- depending on the temperature --- three possible phases: paramagnetic (P), generalized-nematic (N), 
and ferromagnetic (UF). The phase transition between P and N was found to be in the Kosterlitz-Thouless (KT) universality class, while 
the transition between N and UF was found to belong
to the 3-state Potts universality class.  
In his Comment on our work~\cite{korshunov}, Dr. Korshunov argued that the N phase
cannot exist for $q>2$, and that there should only be one ``genuine phase transition'' between the P and UF phases.  
We will now show that the argument of Ref.~\cite{korshunov} is incorrect.

Let us first consider $\Delta=0$. In this case the Hamiltonian
becomes purely $q$-nematic. Changing variables in the partition function, $q \theta_i \rightarrow \bar \theta_i$, 
shows that the model is isomorphic to the usual XY model, but with the low temperature phase N, instead of UF. 
The phase transition from P to N will, therefore, occur at  $T_0 \simeq 0.893$, the same temperature as for 
the standard XY model and will belong to the KT universality class. This, clearly demonstrates that the N phase exists for 
$\Delta=0$. Using Ginibre's inequality~\cite{gin} it is possible to show that the P to N transition will  
also extend to finite $\Delta$~\cite{rom}.  
Furthermore, Ginibre's inequality allows one to derive a rigorous
lower bound~\cite{rom} on the transition temperature between P and N phases,  $T_{\scriptstyle\rm KT}(\Delta) \ge (1-\Delta) T_0$. Since at very low temperature
the system must be in UF phase, this proves the existence of P, N and UF 
phases for small, but finite values of $\Delta$, contradicting the heuristic 
argument of Ref.~\cite{korshunov}.

\begin{figure}[hbt]
\includegraphics[width=8.5cm]{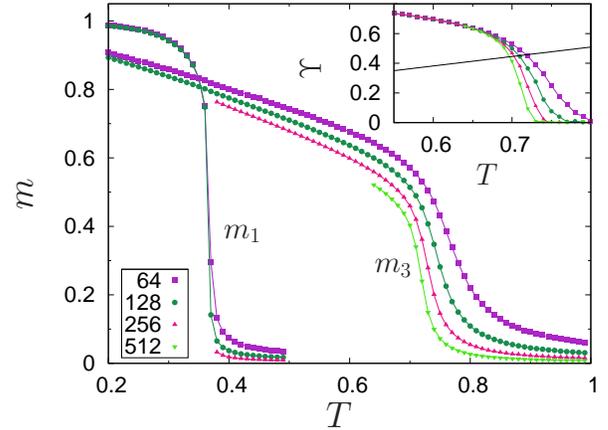}
\caption{Order parameters $m_1$ and $m_3$ (see Ref.~\cite{PoArLe11}) for several system
sizes $L$ showing phase transitions
at $T_{\rm Potts} \simeq 0.365$ and $T_{\scriptstyle\rm KT}\simeq 0.68$.
Inset: Helicity modulus $\Upsilon$ versus $T$. The crossing with the line $2T/\pi$ at
$T_{\scriptstyle\rm KT}(L)$, when extrapolated to $L\to\infty$, 
gives $T_{\scriptstyle\rm KT}\simeq 0.68$.}
\label{fig.helicity}
\end{figure}

To precisely delimit the location of all three  phases for the model with $q=3$,
we consider a specific example, $\Delta=1/4$. For this $\Delta$, and using finite size 
scaling (FSS), in Ref.~\cite{PoArLe11} we have calculated the 
critical temperature for the N-UF transition to be 
$T_{\rm Potts} \simeq 0.365$, which was found to belong
to the 3-state Potts universality class. The order parameter $m_1$
(magnetization) shows clearly this transition, see  Fig.~\ref{fig.helicity}.
On the other hand, at $T_{\scriptstyle\rm KT}$, 
the nematic order parameter $m_3$ shows the transition between N and P phases. 
At the transition temperature, $m_3$ decreases with $L$ as
$m_3(T_{\scriptstyle\rm KT})\sim L^{-\beta/\nu}$, with $\beta/\nu\simeq 0.117$.  
The exponent is very close to the theoretical value expected for
the KT transition, $1/8=0.125$. To further verify the 
``genuineness" of this transition, we calculated the
helicity modulus $\Upsilon$~\cite{fisher}, shown in the inset of Fig.~\ref{fig.helicity} as a function of 
temperature, for several system sizes. The helicity modulus crosses the  straight line $2T/\pi$~\cite{nelson} 
at $T_{\scriptstyle\rm KT}(L)$ and, extrapolating to $L\to\infty$, we obtain $T_{\scriptstyle\rm KT}\simeq 0.68$, slightly above the lower bound provided by Ginibre's inequality.

\begin{figure}[htb]
\includegraphics[width=8.5cm]{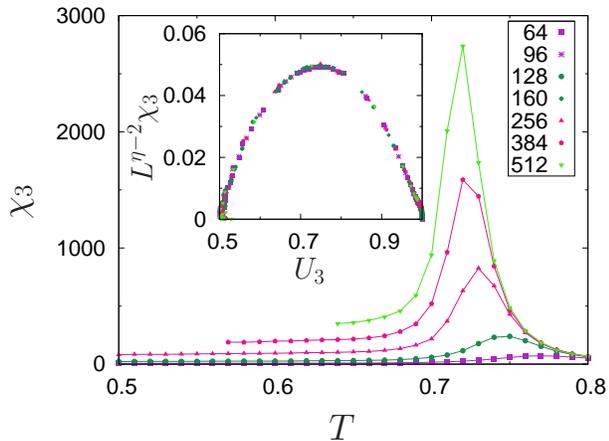}
\caption{The susceptibility $\chi_3$ associated with $m_3$ near the KT transition
for various system sizes. 
Inset: rescaled susceptibility {\it versus} the Binder cumulant, 
$U_3=\langle m_3^2\rangle^2/\langle m_3^4\rangle$, 
showing a perfect collapse with the KT exponent $\eta=1/4$. A similar collapse is also obtained for the
magnetization~\cite{canova}.}
\label{fig.susc}
\end{figure}

In Fig.~\ref{fig.susc} we present the susceptibility $\chi_3$ as a function of $T$
for different system sizes. The phase transition is very clear from the
divergence of the  
susceptibility at $T_{\scriptstyle\rm KT}$, as $L \rightarrow \infty$.
For a KT phase transition, the FSS predicts that 
$\chi_3(T_{\scriptstyle\rm KT})\sim L^{1.75}$,  while our simulations 
find $L^{-1.766}$.  Finally, if we plot $\chi_3 L^{\eta-2}$, with the KT $\eta=1/4$, vs. the Binder cumulant,
all the susceptibilities for different system sizes should collapse onto a universal curve~\cite{loison}.  This is precisely what is found in our MC simulations, see 
inset of Fig.~\ref{fig.susc}. 

Ref.~\cite{korshunov}
also questions the transition between the phases F$_1$ and UF, in the
model with $q=8$, and the absence of any ``qualitative'' difference between these two phases.  
This, however, is clearly not an issue, as is exemplified by the usual liquid-gas phase transition  ---
the difference between liquid and gas being only ``quantitative". The Fig.~6 of
Ref.~\cite{PoArLe11} shows clearly the transition between F$_1$ and UF, which again
belongs to the KT universality class.

In conclusion, we have presented a rigorous proof, as well as, numerical
evidence for the existence of a transition between the N and P phases belonging to 
the KT universality
class, at odds with the heuristic argument of Ref.~\cite{korshunov}.

\end{document}